# Trends and Perspectives for Signal Processing in Consumer Audio

Joshua Atkins and Daniele Giacobello
Beats Electronics, LLC

**Audio On-the-Go: the Headphones Era**

With the advent of streaming music and video services and the increasing miniaturization of electronics over the past decade, a shift has occurred in the way most people discover and consume media. More content is viewed and listened to over headphones and in on-the-go situations than ever before, which has led to both a massive increase in headphone sales and new categories of increasingly small portable speakers. The NPD Group, a U.S. marketing data analytics firm, says the following of the 2013 audio landscape [1]:

> Most audio products continued their positive sales momentum in addition to the stellar headphone results. Sound bar sales increased nearly two times in both unit and dollar volumes, and topped $200 million in revenue for the first time. Streaming audio speakers were the star of the market this year, as revenue increased to $310 million, up 2.5 times from last year.

The headphone sales in 2013 were considered one of the top 5 revenue categories in consumer electronics with estimated U.S. revenue above $500 million (the other 4 were: TVs, tablets, notebooks, and desktops) [2]. This transition speaks not only to the incredible success of perceptual coding leading to the miniaturization of media file sizes, but also to advancements in audio processing for headphone and loudspeaker protection and correction that now run in very low-power devices.

The signal processing for loudspeaker correction and protection traditionally relies on a linear equalization followed by a non-linear compression. This system can be used to constrain the voltage applied to a speaker in order to prevent mechanical excursion beyond a safe handling region [3, 4]. The challenge comes when very small speakers, such as those in cell phones, tablets, and laptops, are used to reproduce full range audio at higher sound pressure levels, as the acoustic output at low frequencies is inefficient. Algorithms for nonlinear system identification and correction have been developed to push these small speakers to the limits of their physical ability [3, 5]. However, they require specialized hardware to provide feedback from the loudspeaker through voltage and current measurements leading sometimes to sophisticated models, which are too complex to implement in real-time systems. A third avenue which has been investigated attempts to recreate the perception of low frequency sounds by analyzing a signal and generating harmonics of low frequency content that the loudspeaker cannot reproduce; the harmonics, even without the fundamental, will still lead to the original low tone being perceived [6]. The challenge with many of these techniques lies in the tradeoff between various audible artifacts and performance, as many require hand tuning by an experienced audio engineer due to the lack of robust objective models for perception of audio quality.

The transition to portable listening has led to a boom in headphone sales with the NPD Group estimating over 95% of the growth happening in the premium ($100+) category [2]. When surveyed, customers typically rank the inclusion of noise cancellation via signal processing (either analog or digital) as a major important factor in their choice of headphone, second only to sound quality [2]. Opportunities for signal processing research lie in the improvement of methods for active control and adaptive filtering applied to noise cancellation and the presentation of three-dimensional spatial audio.

Headphones and earphones deliver sound to the listener in a way unlike loudspeakers as they lack the natural crosstalk between channels and do not present a shift in spatial cues during head movement. However, through signal processing, the acoustic signals that the ear drum would have received in the loudspeaker listening scenario can be recreated by measuring the appropriate impulse responses and running the convolution in real time. With head-tracking and individual measurements, the effect can be very convincing, but challenges arise in generalization and in systems with multiple sources and reverberation where the impulse responses are thousands of taps at normal audio rates [7]. These virtual playback techniques have the possibility of presenting users with more natural spatial content than could be possible in the traditional home theater, as the cost of adding a "virtual speaker" is only in DSP horsepower and not in the purchase of a physical loudspeaker and amplifier. As headphones enter into pervasive use in society, applications for augmented reality using the aforementioned methods for virtual sound source playback along with sophisticated auditory scene analysis mechanisms offer interesting avenues for future multimodal interaction and auditory displays.

**Home Audio: Redefining the High-End Experience**

The home market, while currently not as large as that of headphones, is increasing rapidly with the NPD Group showing 2x and 2.5x sales increases in sound bar and streaming audio speakers between 2012 and 2013, respectively [1]. With sound bars becoming the dominant form of home surround presentation, the challenge of reproducing the perception of audio sources where speakers do not exist (e.g. behind, above, and to the sides of listeners) becomes the major signal processing task. While methods such as wave-field synthesis and cross-talk cancellation have made their way into some products, the signal processing for many such methods is analytic or purely numerical in motivation, and very little work has been done to analyze and compare the perceptual implications of various methods [8, 9, 10].

As the film industry moves towards surround formats with object-based coding, where sounds are not mixed down to the conventional 5.1 or 7.1 speaker configuration, the challenges will become even more complex [11]. Many films have already been mixed and shown in theaters with large loudspeaker arrays using the Dolby Atmos system. Some have won major accolades for their sound alone, e.g., Gravity, but the technology has not transitioned into the home environment due to the bandwidth and number of speakers necessary [12].

Two further challenges in the home environment include methods for automatic room correction, where a signal processing system attempts to identify the sound field in the room and automatically adjust equalization to improve the fidelity at the listening position, and systems for streaming audio that maintain fidelity and synchronization comparable to wired counterparts [13, 14]. In particular, the resurgence of casual home listening provides new challenges for dealing with distributed network of small loudspeakers spread throughout a room [15]. The difficulty in this case is not only on the acoustic side (e.g., distributed signal processing), but in development of new protocols that can advance beyond streaming audio over Bluetooth and AirPlay with better synchronized broadcast capability and quality of service [16].

On the opposite end, work on the recording and transmission of spatial audio has seen great advances recently with the advent of spherical microphone arrays and advances in perceptual audio coding [17, 18, 19]. The promise of capturing the three-dimensional sound field at a sporting event, concert, or other live performance and transmitting it to the home for virtual reproduction could transform the entertainment industry. Already, the prospect of similar ideas for video with devices such as the Oculus Rift, a consumer head-worn virtual reality display, are encouraging film-makers to shoot in 3D and game developers to imagine incredibly immersive new forms of gameplay [20]. As such devices blur the lines between virtual and real, the field of augmented reality for audio and video is poised to see many new applications not only in the consumer world, but in areas such as medicine, sports training, and business collaboration.

**Smart Devices: Interaction and Awareness**

The human interaction paradigm with audio rendering devices has also seen a dramatic shift, as devices get smaller and more portable. In the home environment, well-established interaction media such as remote controls are often no longer adequate given the deluge of content that a user can access [21, 22]. Smart interfaces based on automatic speech recognition offer a natural solution to this problem given the hands-busy, mobility-required scenarios, where some of these devices might also be used [23, 24].

A key aspect to obtain a seamless integration of these devices is for them to become increasingly aware of their surrounding environment, i.e., being able to understand the mood of their environment and react accordingly [25]. Devices that infer the type of conversation a group of people are having by speaker recognition, turn-taking behavior analysis, emotion recognition, and keyword spotting can enhance the quality of experience of the user by, for instance, automatically choosing audio content or simply increasing or lowering volumes to appropriate levels [26, 27].

Furthermore, the music content itself is no longer restricted to what users actually own on their playback device. Through connection to music streaming services like Spotify, Pandora, and Beats Music, which track information regarding the music preferences of the user, these devices can use music information retrieval to find new content the user might like, as well as match the environment and mood to a given track or playlist [28, 29].

The integration of these devices into everyday life provides interesting new topics of research, linking problems in audio and acoustics with problems in speech and language processing, as well as general machine learning and music information retrieval. Moreover, the prospect of allowing high quality video calls through these devices embedded with several microphones offers enormous opportunities of research for audio and acoustic signal processing engineers.

The first and foremost problem is audio acquisition and enhancement: microphones in consumer devices are often embedded within the device and very close to the speakers, with a signal to echo ratio that can be extremely low [30]. A solution seems to come from wireless microphone networks which would be able to exploit multiple microphones located in different devices, for example, streaming devices not in use or microphones embedded in various electronics in order to cover a larger area of interest [31]. The inherent latency embedded within each node, however, does not seem to allow a real-time solution in the near future with the current streaming codecs and protocols. In the short- to medium-term, microphone arrays seem to be the only commercially viable solution [32, 33]. However, new solutions based on data driven approaches [34, 35], like source separation [36] and auditory scene analysis [37], and other methods that can incorporate more than loose statistical assumptions [38, 39, 40], as well as robust estimation methods [41, 42], can pave the way to more robust solutions [43].

Furthermore, new advances in embedded convex optimization will finally unleash a brand new set of algorithms and software, unfeasible for real-time applications only a couple of years ago [44, 45]. Benefits of the increasing awareness of the devices can extend beyond the entertainment context into health applications such as assistive living technologies, creating smart environment and intelligent companions that can preserve independence and security of people with special needs [46, 47].

**Conferences and Special Sessions on Audio and Acoustic Signal Processing**

For those interested in more in-depth technical details, we would like to bring attention to these upcoming events addressing some of the challenges presented above which might be of interest to the community:

- *International Conference on Auditory Displays*, New York, NY, June 22-25, 2014
- *17th International Conference on Digital Audio Effects*, Erlangen, Germany, September 1-5, 2014
- *137th International Convention of the Audio Engineering Society*, Los Angeles, California, October 9-12, 2014
- *55th Audio Engineering Society International Conference on Spatial Audio*, Helsinki, Finland, August 27-29, 2014
- *Special Session on Digital Audio Processing for Loudspeakers and Headphones, EUSIPCO*, Lisbon, Portugal, September 1-5, 2014 (organized by the authors)
- *IEEE Journal of Selected Topics in Signal Processing - Special Issue on Spatial Audio*, Manuscript submission: July 1, 2014
- *IEEE Journal on Selected Topics in Signal Processing - Special Issue on Interactive Media Processing for Immersive Communication*, Manuscript submission: April 16, 2014
- *Machine Learning Applications in Speech Processing, GlobalSIP*, Atlanta, Georgia, December 3-5, 2014
- *Special Sessions at INTERSPEECH*, Singapore, 14-18 September 2014
    - A Re-evaluation of Robustness
    - Multichannel Processing for Distant Speech Recognition
    - Speech technologies for Ambient Assisted Living